\documentstyle[prl,aps,preprint,tighten,floats]{revtex}
\def\simlt{\stackrel{<}{{}_\sim}}
\def\simgt{\stackrel{>}{{}_\sim}}
\begin{document}
\draft
\preprint{UPR-0790-T, hep-ph/9805281}
\date{\today}
\title{A Mechanism for Ordinary-Sterile Neutrino Mixing} 
\author{Paul Langacker}
\address{Department of Physics and Astronomy \\ 
          University of Pennsylvania, Philadelphia PA 19104-6396\\ }
\maketitle

\begin{abstract}
Efficient oscillations between ordinary (active) and sterile
neutrinos can occur only if Dirac and Majorana mass
terms exist which are both small and comparable. It is shown that
this can occur naturally in a class of string models, in which
higher-dimensional operators in the superpotential lead 
to an intermediate scale expectation value for a scalar field
and to suppressed Dirac and Majorana fermion masses.
\end{abstract}
\pacs{}

\section{Introduction}

There have been suggestions \cite{hints} 
that there may be mixing between
ordinary neutrinos (with normal weak interactions) and sterile neutrinos
(which have no ordinary interactions except by mixing) \cite{sterile}. 
This is motivated by the various
hints for neutrino masses or oscillations, including the MSW interpretation
of the solar neutrino spectral anomaly \cite{MSW,MSWnew}, the anomalous $\mu/e$ ratio
produced by atmospheric neutrinos \cite{atmospheric}, 
the candidate events for
$\nu_\mu \rightarrow \nu_e$ (or $\bar{\nu}_\mu \rightarrow \bar{\nu}_e$)
in the LSND experiment \cite{LSND}, and the possibility of a hot (neutrino)  
component to the dark matter (mixed dark matter) \cite{mixed}. 
The first three of these suggest
neutrino oscillations, most likely with different ranges for the
neutrino mass-squared differences $\Delta m^2_{ij} \equiv m_i^2 - m_j^2$,
where $m_i$ is the mass of the $i^{th}$ mass eigenstate neutrino, 
around $10^{-5}$, 
$10^{-3}-10^{-2}$, and $10^{-1} - 10$ eV$^2$, respectively\footnote{Several
authors \cite{threenu} have suggested that it might be possible to accomodate
the atmospheric neutrinos and LSND by a single $\Delta m^2 \sim 10^{-1}$
 eV$^2$ in a three-neutrino mixing scheme. However, this range is strongly
disfavored by the azimuthal and energy distributions in the
most recent Superkamiokande data \cite{superK}.}. The mixed dark matter 
scenario suggests one or more mass eigenstate neutrinos $\nu_i$
in the several eV 
range. Given the constraint 
$N_\nu = 2.990 \pm 0.011$ from the $Z$ lineshape
\cite{lineshape} on the number of light ordinary neutrinos, 
one cannot accomodate all of
these possibilities unless there are additional
sterile neutrinos (which don't contribute significantly to $N_\nu$)  
\cite{hints}.
Mixed dark matter also suggests the possibility of
near degeneracies, so that $|\Delta m^2_{i,j}| \ll m_{i,j}^2$ \cite{hints}. 

Another implication is for big bang nucleosynthesis \cite{BBN}. There are
conflicting estimates of the abundance of primordial deuterium
from QSO absorption lines, but the low $D$ observations, 
combined with canonical estimates of the primordial $^4He$ abundance,
suggest only $N_{eff} =1.9 \pm 0.3$ light neutrinos in equilibrium at the
time of decoupling of the neutron to proton ratio \cite{crisis}. 
The resolution
may well be that the systematic uncertainties in the $^4He$ abundance
have been underestimated (i.e., that there is more $^4He$, as predicted by
$N_{eff} = 3$), or possibly that the high $D$ QSO observations
are correct. However, it is possible that the number of effective neutrinos
$N_{eff}$ is indeed smaller than three. Mechanisms for achieving this
include a decaying $\tau$ neutrino, which for certain masses, lifetimes,
and decay modes can lead to a lower energy density \cite{tau}. 
Another possibility
is a significant  $\nu_e$ -- $\bar{\nu}_e$ asymmetry \cite{degenerateBBN}. 
A $\nu_e$ 
excess would deplete the $n/p$ ratio prior to the decoupling of the
$\nu_e n \leftrightarrow e^- p$ reaction, leading to the production of less
$^4He$, and thus $N_{eff} < 3$.

The mixing of ordinary and sterile neutrinos could affect $N_{eff}$
in two ways.  The combination of ordinary-sterile oscillations
with the rescattering of the active component (which destroys the
phase coherence of the oscillating state and serves as a measurement)
can lead to the production of sterile neutrinos. Ignoring effects of
lepton asymmetries, a number of authors \cite{sterileBBN} have argued that
for a wide range of ordinary-sterile neutrino parameters, 
the mixing could lead to the production of light sterile neutrinos
in equilibrium numbers prior to nucleosynthesis, 
increasing the energy density and increasing $N_{eff}$ to four, 
aggravating the difficulty. However, Foot and Volkas have recently
argued \cite{FV}
that the asymmetric interaction of neutrinos and antineutrinos
with matter could, for a  significant parameter range, lead to the
generation of a large neutrino-antineutrino asymmetry prior to 
nucleosynthesis. This could both suppress the production of sterile 
neutrinos, and in some cases lead to a significant excess of 
 $\nu_e$ with respect to $\bar{\nu}_e$, thus weaking the nucleosynthesis
constraint on ordinary sterile mixing, or even accounting for
$N_{eff} < 3$.

Yet another motivation for ordinary-sterile neutrino mixing comes from
heavy element synthesis in supernova explosions. One promising site for
the (neutron enriched) $r$-processes are in the ejecta of
neutrino-heated supernova explosions \cite{rprocess}. Unfortunately, 
$\nu_e$ emissions would render the $r$-process impossible, due to
the destruction of neutrons by $\nu_e n \rightarrow e^- p$.
The problem would be even worse for $\nu_e \leftrightarrow \nu_\mu$ or
($\nu_\tau$) conversions with  $\Delta m^2 > 10^{-2}$ eV$^2$, which
would create more energetic $\nu_e$'s \cite{rMSW}. 
However, it has recently been 
argued by Caldwell, Fuller, and Qian \cite{rsterile}
that ordinary-sterile neutrino mixing would
yield a robust solution to the difficulty, by allowing $\nu_\mu$
to convert to sterile neutrinos and escape from the supernova,
followed by $\nu_e$ converting into (harmless) $\nu_\mu$.

Clearly, none of these motivations for ordinary-sterile neutrino
mixing is compelling, but nevertheless they motivate an examination
of the theoretical possibilities for 
significant ordinary-sterile neutrino mixing. 
In Section II I review the basic
constraints on  Dirac and Majorana neutrino masses needed to generate
significant ordinary-sterile mixing. In particular, it will only
occur in theories in which the Dirac and Majorana masses are both
small and comparable to each other \cite{sterile}, 
which is difficult to achieve
without fine-tuning in most models of neutrino mass \cite{review}. 
In Section III,
however, I discuss a framework in which it is quite plausible to
have the necessary ingredients without fine-tuning. It was recently 
argued \cite{cceel} that in perturbative superstring vacua realistic
hierarchies of quark and charged lepton masses may be generated
by higher-dimensional operators, in which the effective Higgs
Yukawa couplings are suppressed by powers of
the ratio of an intermediate scale vacuum expectation value to the
string scale. It was also shown that in such models  
it is possible to have naturally small
Dirac neutrino masses, as well as Majorana masses for the sterile
neutrinos that can be either large or small depending on the dimensions
of the relevant operators. These ideas are extended
in Section III, in which it is shown that the Dirac and Majorana
masses can indeed be small and comparable if two conditions are
satisfied: (i) the scale $m_{soft}$ of supersymmetry breaking in the
observable sector is comparable to the electroweak scale (i.e., no more than
a TeV), as is necessary if supersymmetry is to be relevant to the
stabilization of the electroweak scale and is predicted in models
of radiative electroweak breaking; (ii) a simple relation is satisfied
between the (integer) dimensions of the higher-dimensional operators
responsible for the Dirac masses, the Majorana masses, and an
operator associated with the intermediate scale vacuum expectation
value. Although this relation will not be satisfied in all
or even most
string compactifications, it is sufficiently simple as to provide a
plausible framework for ordinary-sterile mixing.

\section{Ordinary-Sterile Neutrino Mixing}
An ordinary (active) neutrino occurs in an $SU(2)$ doublet with
a charged lepton, and thus has standard charged and neutral current
weak interactions. For example, the left-chiral $\nu_{eL}$ is the
partner of the left-handed $e^-_L$. $\nu_{eL}$ is necessarily 
associated by CPT with the right-chiral antineutrino $\nu^c_{eR}$,
the $SU(2)$ partner of the $e^+_R$. $\nu_{eL}$ and $\nu^c_{eR}$
together constitute a Weyl two-component neutrino.
A sterile neutrino 
is an $SU(2)$ singlet Weyl neutrino\footnote{Which chiral state
is referred to as the particle, and which as the antiparticle,
is mainly a matter of convention, or is motivated if there is a
conserved or approximate lepton number in the theory.}
 $N^c_L$ and its CPT partner
$N_R$. It has no gauge
interactions except by mixing. Most extensions of the
standard model involve such sterile neutrinos. The only
real issues are whether the sterile neutrinos are light, and whether
there is significant mixing between the ordinary and sterile states
of the same chirality.

In the presence of one ordinary and one sterile Weyl neutrino, the
most general mass matrix is \cite{review}
\begin{equation}
 -L = \frac{1}{2} 
( \begin{array}{cc} \nu_L & N^c_L \end{array} )
\left( \begin{array}{cc} m_T & m_D \\ m_D^T & m_M \end{array} \right)
\left( \begin{array}{c} \nu^c_R  \\ N_R \end{array} \right) + h.c.,
\label{matrix}
\end{equation}
where $m_T$ and $m_M$ are  Majorana mass terms for
the ordinary and sterile neutrinos, respectively, and $m_D$ 
is a Dirac mass term, which connects two distinct Weyl neutrinos.
$m_T$, $m_M$, and $m_D$
break weak isospin by 1, 0, and 1/2 units, respectively, and
can be generated by the vacuum expectation values of Higgs triplets,
singlets, and doublets\footnote{$m_M$ could in principle be a bare
mass, but this is  forbidden by additional symmetries in most
extensions of the standard model.}. In many models, $m_T$ is absent.
Diagonalizing (\ref{matrix}), one obtains two  Majorana mass
eigenstate neutrinos, $\nu_1$ and $\nu_2$. The left chiral states are
related by
\begin{eqnarray}
 \nu_L& = &  \nu_{1L} \cos \theta + \nu_{2L} \sin \theta \nonumber \\ 
N^c_L & = & -\nu_{1L} \sin \theta + \nu_{2L} \cos \theta,
\label{mixing}
\end{eqnarray}
where $\theta$ is the mixing angle. A similar relation holds for the
right-chiral components.
In the generalization to three families,
one can interpret $\nu_L$, $N^c_L$, $\nu^c_R$, and $N_R$ as
three-component vectors in family space, and $m_T$, $m_D$, and $m_M$ as
$3 \times 3$ matrices ($m^T_D$ is the transpose of $m_D$; $m_T$ and $m_M$
are symmetric), yielding six 
Majorana mass eigenstates\footnote{For a general discussion, see 
\cite{review}.}. 
The further generalization to models in which 
there are are a different number of sterile states is straightforward.

Most models of neutrino mass involve limiting cases of (\ref{matrix})
in which there is little or
no mixing between $\nu_L$ and 
$N^c_L$. For example, the pure Dirac case ($m_T = m_M = 0$)
yields two degenerate Majorana eigenstates, which can be combined to form
a four-component Dirac neutrino. One finds $\theta = \pi/4$, but  can
transform to  an appropriate basis for the degenerate states in which it
is manifest that there is a conserved lepton number and there are no  
$\nu_L \rightarrow N^c_L$ transitions. In the pure Majorana
case\footnote{The simplest model \cite{gr} with $m_T \ne 0$ is 
excluded phenomenologically \cite{review}.}, $m_D$ = 0, one has
$\theta = 0$, and the sterile neutrino decouples. 
Similarly, in the seesaw limit\footnote{One usually
assumes $m_T = 0$ in the seesaw model, but similar conclusions apply
if $m_T$ is comparable to $m_D^2/m_M$.} \cite{seesaw}, 
$m_M \gg m_D$, there is
one naturally light Majorana state $\nu_{1L}  \sim \nu_L$ with
$m_1 \sim m_D^2/m_M$, and one heavy state $\nu_{2L}  \sim N^c_L$ with
$m_2 \sim m_M$, with negligible mixing ($\theta \sim m_D/m_M$).

Significant ordinary-sterile mixing can only occur if $m_D$
is of the same order of magnitude as $m_M$ and/or $m_T$. For definiteness,
I will concentrate on $m_M$. Because
of the limits on neutrino mass, one needs not only that $m_D$ and
$m_M$ are comparable, but both must be much smaller than the quark and
charged lepton masses. If these conditions are satisfied, then
ordinary-sterile (referred to as second class \cite{sterile})
neutrino oscillations
between $\nu_L$ and $N^c_L$ can occur, with the usual formula
\begin{equation}
P(\nu_L \rightarrow N^c_L) = \sin^2 2 \theta \sin^2 \left(
\frac{1.27 \Delta m^2 (eV^2) L (km)}{E (GeV)} \right),
\label{oscillation}
\end{equation}
where $\Delta m^2 = m^2_2 - m^2_1$, $L$ is the distance traveled,
and $E$ is the energy.

Similar statements apply to the $6 \times 6$ generalization. 
In the pure Dirac, pure Majorana, or seesaw limits, there
can be ordinary flavor (first class) oscillations between the three
ordinary neutrinos, but no second class oscillations between the
$\nu_L$ and the $N^c_L$ because of the absence of significant
mixing between the two sectors (and because the sterile states are
heavy in the seesaw model). However, if the Dirac and Majorana masses
are small and comparable, then there can be signficant mixing between
the six left-chiral states, and both first and second class oscillations
can occur.

Thus, significant ordinary-sterile mixing can occur, but it appears
to require two miracles, i.e., small Dirac and Majorana masses. However, in
the next Section I will describe a plausible framework in which this
can occur naturally\footnote{Another possible mechanism is described
in \cite{othermech}.}.

\section{Mass Generation by Higher-Dimensional Operators}

Recently, a simple mechanism was found to generate small Dirac neutrino
masses without invoking a seesaw, and also Majorana masses for the
sterile neutrinos that could be either large or small \cite{cceel}.
It utilized higher-dimensional operators\footnote{Other studies of the
implications of higher-dimensional operators for neutrino masses
may be found in \cite{otherhdo}.}  in the
superpotentials of perturbative superstring compactifications 
based on the free fermionic construction \cite{freefermion} with
additional non-anomalous $U(1)$ gauge factors. The basic
idea is that these operators involve additional fields which are singlets
under the standard model gauge group. In some cases, these fields acquire
intermediate scale vacuum expectation values, 
leading to effective Yukawa couplings 
that are suppressed by powers of the ratio of the intermediate scale to the
string scale. Here, I will show that in some cases the Dirac and Majorana
masses will naturally be small and comparable, although it is hard
to make precise numerical predictions without a specific fully 
realistic string model. Some of the considerations could also
occur in a more general class of models, but they are 
especially well motivated
in the case of the superstring models with an extra $U(1)$.

In the typical supergravity model \cite{sugra} with radiative electroweak
breaking, the  mass-squares of the Higgs doublets $H_{1,2}$
at the Planck scale are positive
and of order $m_{soft}^2$, where $m_{soft}$ is a typical soft
supersymmetry breaking parameter. However, the large 
top-Yukawa drives one of the running Higgs mass-squares to a negative
value of order $-m_{soft}^2$ at low energy, leading to
electroweak breaking at a scale $\langle H_2 \rangle \sim m_{soft}$.
Many superstring models involve
an extra non-anomalous $U(1)$ in the observable sector.
Recently, it was argued \cite{cl} that 
this $U(1)$ would either: (i) be unbroken; (ii) be broken near
the electroweak scale (i.e., below 1 TeV); or (iii) broken
at a scale intermediate between the electroweak and string
scales. This is because in most string models with
supergravity mediated supersymmetry breaking  all of
the scalar mass-squares are positive and of the same order of magnitude,
$m_{soft}^2$, at the string scale. (These string models do not allow
supersymmetric mass terms.)
Furthermore, Yukawa couplings at
the string scale are typically either zero or simply related to a
gauge coupling of order unity. Breaking can occur if there is a  standard
model singlet field $S$ that is charged under the extra $U(1)$ and
which has a large Yukawa coupling to other fields.
Then a radiative mechanism can occur
analogous to the radiative electroweak breaking, i.e., the
scalar would  acquire an expectation value
$\langle S \rangle \sim m_{soft}$, breaking the $U(1)$ near the electroweak
scale\footnote{Such models were more fully explored and shown to yield a
natural solution to the $\mu$ problem in \cite{cdeel}.}. 

However, if there are two or more such fields 
with opposite signs for their
$U(1)$ charges, then, depending on the details of the soft breaking, 
the minimum may occur along a $D$-flat direction. In many string models,
these directions are also $F$-flat up to higher-dimensional terms.
For example, 
for two fields $S_{1,2}$ with equal and opposite charges $Q$,
the low energy potential is
\begin{equation}
V(S_1,S_2) = m_1^2 |S_1^2| + m_2^2 |S_2^2| + \frac{g'^2 Q^2}{2}
(|S_1^2| - |S_2^2|)^2,
\label{potential}
\end{equation}
where $g'$ is the gauge coupling of the extra $U(1)$. For
$m_1^2 < 0$ but
$m_1^2 + m_2^2 > 0$, the minimum occurs for
$\langle S_1 \rangle \sim m_{soft}$ and $\langle S_2 \rangle = 0$,
similar to the single field case. However, for $m_1^2 + m_2^2 < 0$,
the minimum occurs along the $D$-flat direction\footnote{A similar 
situation could 
occur for a scalar field $S$ not charged under any gauge group.}
$\langle S_1 \rangle = \langle S_2 \rangle \equiv \langle S \rangle$,
for which
\begin{equation}
V(S) = m^2 |S^2|,
\label{dflat}
\end{equation}
where $m^2 = m_1^2 + m_2^2$. This potential appears to be unbounded from below.
However, the potential will be stabilized by one or both of the
following mechanisms: (i) higher-loop terms in the effective potential
lead to the replacement of $m^2$ in (\ref{dflat}) by the running $m^2(S)$,
evaluated at the scale $S$. This renormalization group improved potential
has a minimum close to the point $\mu_{RAD}$ at which $m^2$ crosses
zero \cite{cceel}. $\mu_{RAD}$ is sensitive to the soft parameters and
Yukawas, and can occur anywhere between 1 TeV and the string scale.
(ii) Higher-dimensional terms in the superpotential involving $S$, which
are generally expected in string models, can also stabilize the potential.
The latter mechanism usually dominates whenever $\mu_{RAD} \simgt 10^{12}$
GeV. The consequences of both mechanisms were explored in \cite{cceel}.
In particular, it was shown that other higher-dimensional terms involving
$S$ could also lead to a reasonable effective $\mu$ parameter, and to
interesting hierarchies of quark, charged lepton, and neutrino masses. 

For definiteness, I will consider the case in which the minimum of the
potential occurs along a one-dimensional flat direction, characterized
by an effective field $S = s/\sqrt{2}$. The potential for $s$ is \cite{cceel}
\begin{equation}
\label{potflat}
V(s)= \frac{1}{2}m^2s^2+\frac{1}{2(K+2)}
\left(\frac{s^{2+K}}{{\cal M}^K}\right)^2,
\end{equation}
where ${\cal M}$ is of the order of the string scale (e.g., ${\cal M}
\sim 10^{17}-10^{18}$
GeV), and $m^2 = {\cal O} (-m_{soft}^2)$ is the running soft mass squared
evaluated at $s$. The second term, with $K=1,2,\cdots,$
is a higher-dimensional
operator associated with the non-renormalizable term
\begin{equation}
\label{nrsup}
W_{\rm NR}=\left(\frac{\alpha_{K}}{M_{Pl}}\right)^{K}\hat{S}^{3+K},
\end{equation}
in the superpotential. In (\ref{nrsup}), 
$\hat{S}$ is the superfield corresponding
to $S$, $M_{Pl}$ is the Planck scale, and
$\alpha_{K}$ is a calculable coefficient in a given string model.
${\cal M}$ and $M_{Pl}$ are related by
${\cal M} = {C}_K M_{Pl}/ \alpha_K$, where
${C}_K = [ 2^{K+1}/((K+2)(K+3)^2)]^{1/(2K)}$.
I will assume that the potential is stabilized by the
higher-dimensional operator and that the running of $m^2$ can be
neglected for the minimization. Then, the minimum occurs
for
\begin{equation}
\label{vev}
\langle s\rangle=
\left[\sqrt{(-m^2)}{\cal M}^K \right]^\frac{1}{K+1}
\sim  (m_{soft}{\cal M}^K)^\frac{1}{K+1}.
\end{equation}
As was discussed in \cite{cceel}, the electroweak scale in the
same model can occur by the radiative mechanism at
$\langle H_{1,2} \rangle = {\cal O} (m_{soft})$.

Dirac neutrino masses may be generated by the operators \cite{cceel}
\begin{equation}
W_{\rm D} \sim \hat{H}_2 \hat{L}_L \hat{N}^c_L 
            \left( {\hat{S}\over {\cal M}} \right)^{P_D},
\label{Diracnu}
\end{equation}
where $\hat{L}_L$ is the superfield corresponding to a left chiral
doublet which includes an active $\nu_L$, and $\hat{N}^c_L$ is
the superfield corresponding to a sterile neutrino. This yields
an effective Dirac Yukawa coupling of order $(\langle S 
\rangle/{\cal M})^{P_D}$.
Similar terms can lead to a
reasonable hierarchy of quark and charged lepton masses. The integer
powers $P_D$ can differ for quarks, charged leptons, and neutrinos,
and can be generalized to include family indices and generational mixing.
They are model dependent, but calculable in a specific string model.
In some cases, the allowed powers for each type of term are restricted
by the $U(1)$ and other symmetries of the effective field theory.
However, it is often the case that terms that would be allowed by the
effective field theory symmetries are absent due to selection
rules in the underlying string theory\footnote{For a recent discussion, see
\cite{cceel2}.}. It is reasonable that in some models the neutrino
mass terms occur in higher orders than those for the quarks and 
charged leptons, leading to naturally small Dirac neutrino masses.

Majorana mass terms for both the ordinary and active neutrinos may
be generated by the higher-dimensional operators
\begin{equation}
W_{\rm T} \sim 
              \frac{\left(\hat{H}_2 \hat{L}_L \right)^2}{{\cal M}}
              \left( {\hat{S}\over {\cal M}} \right)^{P_T}
\label{tripletmajorana}
\end{equation}
and
\begin{equation}
W_{\rm M} \sim \hat{N}^c_L \hat{N}^c_L \hat{S}
\left( {\hat{S}\over {\cal M}} \right)^{P_M},
\label{singletmajorana}
\end{equation}
respectively.

Hence, one obtains neutrino masses
\begin{eqnarray}
 m_D & \sim & \langle H_2\rangle \left( \frac{\langle S \rangle}{\cal M}
\right)^{P_D} \ \, \sim
\langle H_2\rangle
\left(\frac{m_{soft}}{\cal M}\right)^{\frac{P_D}{K+1}},
\nonumber \\ 
 m_T & \sim & \frac{\langle H_2\rangle^2}{\cal M} 
\left( \frac{\langle S \rangle}{\cal M}
\right)^{P_T} \sim 
\frac{\langle H_2\rangle^2}{\cal M} 
\left(\frac{m_{soft}}{\cal M}\right)^{\frac{P_T}{K+1}}, \nonumber \\ 
 m_M & \sim & \langle S \rangle
\left( \frac{\langle S \rangle}{\cal M}
\right)^{P_M} \ \ \ \,  \sim
 m_{soft}
\left(\frac{m_{soft}}{\cal M}\right)^{\frac{ P_M-K}{K+1}},
\label{nuvalues}
\end{eqnarray}
which depend on the powers $K$, $P_D$, $P_T$, and $P_M$.
The second expressions in (\ref{nuvalues}) result from substituting
(\ref{vev}) for  $\langle S \rangle$.

One expects $ m_{soft}/{\cal M} \sim 10^{-14}-10^{-16}$ for
$ m_{soft} \sim 100 - 1000$ GeV and ${\cal M} \sim 10^{17}-10^{18}$
GeV. Since $\langle H_2\rangle = {\cal O}$(100 GeV),
$m_D$ can be in the interesting range of $10^{-3} - 10$ eV 
for ${P_D}/(K+1) \simlt 1$. One has $\langle H_2\rangle^2/{\cal M}  
\sim 10^{-4}$ eV for ${\cal M} \sim 10^{17}$ GeV, so $m_T$ is
too small to be phenomenologically interesting unless $P_T = 0$.
Even then, $m_T$ is too small to be relevant to the MSW conversions
of solar neutrinos unless one stretches the parameters \cite{MSW}.
It could possibly be relevant to vacuum oscillations  \cite{vacuum}. 

The sterile neutrino Majorana mass $m_M$
may be large or small.
For $P_M - K < 0$ one has $m_M \gg m_{soft}$, allowing a conventional
seesaw model. However, $m_M$ is naturally small for $P_M - K > 0$.
The special case $P_D = P_M - K$  is particularly interesting,
because it implies $m_M/m_D \sim m_{soft}/\langle H_2\rangle
= {\cal O}(1)$, just the condition needed for significant
ordinary-sterile neutrino mixing. This is quite encouraging for
this class of models. However, it is difficult to be more 
quantitative because the precise relation between $m_{soft}$
and $\langle H_2\rangle$, and the values of the effective
mass ${\cal M}$, which could vary somewhat from operator to
operator, are model dependent.

\section{Conclusions}

There are a number of laboratory, astrophysical, and cosmological
hints for neutrino masses and mixing. While it is too early to
be certain of any or all of these, the pattern suggests the possibility
of significant ordinary ($SU(2)$-doublet) -sterile ($SU(2)$-singlet)
neutrino mixing. Such mixing could
even play an important role in understanding 
heavy element synthesis in supernova explosions. It is therefore useful
to seriously examine the theoretical possibilities for ordinary-sterile
mixing.

Most extensions of the standard model predict the existence of sterile
neutrinos. The only real questions are whether the ordinary and
sterile neutrinos of the same chirality mix significantly with
each other, and whether the mass eigenstate neutrinos are sufficiently
light. When there are only Dirac masses, the ordinary and sterile
states do not mix because of the conserved lepton number. Pure
Majorana masses do not mix the ordinary and sterile sectors either.
In the seesaw model the mixing is negligibly small, and the 
(mainly) sterile eigenstates are too heavy to be relevant to oscillations.
The only way to have significant mixing and  small mass eigenstates
is for the Dirac and Majorana neutrino mass terms to be extremely
small and to also be comparable to each
other\footnote{I am using the term Dirac to refer to a mass term 
connecting a left-chiral doublet with a right-chiral singlet,
and Majorana for a term connecting either left and right-chiral singlets
or left and right-chiral doublets. In generalizations involving two or more
singlet Weyl neutrinos there are limiting cases in which pairs of Majorana
mass eigenstates are degenerate and can be combined to form a
four component (Dirac) neutrino  with a conserved lepton number. Whether
one chooses to refer to these masses as Dirac is a matter of convention.
There is an analogous situation for pairs of doublet Weyl neutrinos
(the Mahmoud-Konopinski model). Even in these cases the relevant
singlet-singlet or doublet-doublet mass terms are still controlled by
operators analogous  to (11) and (10), respectively. The
estimates in (12) and implications for ordinary-sterile mixing still apply.
I am grateful to Y. Grossman for emphasizing this case to me.}.
This appears to
require two miracles in conventional models of neutrino mass.

However, it has been argued in this paper that there is a natural
framework for achieving the necessary conditions. The essential
ingredient is that the effective Yukawa couplings which induce the
Dirac and Majorana masses are suppressed by integer powers of
$\langle S \rangle/{\cal M}$, where $S$ is a standard model singlet 
which can acquire an expectation value $\langle S \rangle$ at 
a scale intermediate between the electroweak (or supersymmetry-breaking)
scale and a large (e.g., string) scale ${\cal M}$. 
This situation is motivated by
perturbative superstring compactifications, which often have
additional non-anomalous $U(1)$ gauge symmetries in the observable
sector. These may lead to intermediate scale breaking if the
minimum of the scalar potential occurs along a $D$-flat direction
\cite{cceel}. Furthermore, such theories involve higher-dimensional
operators needed to stabilize the scalar potential, generate
an effective electroweak $\mu$ term, and generate effective
Yukawa couplings for hierarchies of quark, charged lepton, and
neutrino masses. The allowed higher-dimensional operators depend
in general on both the symmetries of the effective four-dimensional
field theory and on the underlying string dynamics.

For appropriate higher-dimensional operators, it is possible to achieve
small Dirac neutrino masses (without invoking a seesaw), and Majorana
masses which can be either large or small. When a specific condition,
$P_D = P_M - K$,  for the dimensions of the operators defined
in (\ref{nrsup}), (\ref{Diracnu}), and (\ref{singletmajorana}) is
satisfied, then the Dirac and Majorana masses will indeed be comparable
for a supersymmetry breaking scale less than around 1 TeV, leading
to significant mixing. This condition is very specific and is
not expected to hold in all or even most
compactifications, so ordinary-sterile mixing
cannot be considered to be a prediction of this class of models.
It is nevertheless a
simple relation between (presumably small) integers which could
hold in many models. It is certainly not a case of fine-tuning.

It is difficult to be more precise than the order of magnitude
estimates presented in this paper. The exact relation between
the soft breaking scale and the expectation values of $S$ and
the Higgs doublets depends on the details of the soft breaking 
and the running of the parameters to low energies. Also,
the coefficients of the various higher-dimensional operators
(the $\alpha_{K}$ in (\ref{nrsup})) are model dependent. Detailed
calculations would require a specific realistic string model and
supersymmetry breaking mechanism. However, the basic ideas 
are expected to hold in a wide class of models.

The framework presented here may provide a viable mechanism for
small Dirac and Majorana masses and significant ordinary-sterile
mixing. Unfortunately, if will be difficult to establish such
mixing unambiguously. Ordinary-sterile oscillations are probably only
observable in disappearance experiments. In principle,  the sterile
neutrinos could be observed directly, e.g., by the production of the
wrong sign lepton in appearance experiments, if they
have (suppressed) interactions from new physics, but such effects are
expected to be extremely small \cite{wang}. If ordinary-sterile mixing
really occurs, it will be necesssary to establish it by difficult
indirect methods, such as careful measurements of the spectral
distortions in solar neutrino experiments \cite{solarspectrum},
or by angular distributions \cite{atmangle} or neutral current
rates \cite{atmnc} for atmospheric neutrinos.

\acknowledgments

This work was supported in part by U.S. Department of Energy Grant No. 
DOE-EY-76-02-3071. It is a pleasure  to thank M. Cveti\v c, G. Cleaver,
J. R. Espinosa, L. Everett, and J. Wang for collaborations on works
closely related to this paper, and V. Barger and
D. Caldwell for useful discussions.

\end{document}